# ANONYMITY AND VERIFIABILITY IN MULTI-ATTRIBUTE REVERSE AUCTION

T. R. Srinath[1], Mahendra Pratap Singh[1] and Alwyn Roshan Pais[1]

[1] Information Security Lab, Dept. of Computer Science and Engineering,
NITK-Surathkal, India-575025
`talari.srinath@gmail.com, mahoo15@gmail.com and alwyn.pais@gmail.com`

## ABSTRACT

*The use of e-Auction services has been increasing in recent years. Security requirements in conducting e-Auctions are mainly bid privacy, anonymity and public verifiability. Most of the secure protocols concentrate on privacy and anonymity, which are achieved through bidder-resolved multi-party computation, assuming two or more trusted third parties, either through numerous auctioneers or with asymmetric models in which the commercial entity of an auction issuer or registration manager is assumed in addition to the auctioneer. Multi-attribute reverse auctions promise higher market efficiency and effective information exchange. This work extends and uses the existing schemes. This scheme uses scoring function, winner determination in multi-attribute auctions to implement public verifiability. Anonymity is achieved through bidder side pseudonym generation. By results and analysis we say this is very simple and effective scheme. This scheme ensures public verifiability and anonymity in multi-attribute auctions without revelation of the bids received, third parties and complex communications.*

## 1. INTRODUCTION

E-auction is an efficient method to buy and sell the items on the Internet. An auction is a mechanism based on a pair of rules, namely the allocation rule that defines which good is allocated to whom and the payment rule that defines the charge of the auction winner(s) [1]. A participant of an auction is called bidder, while the entity conducting the auction is auctioneer. Today, auction is a popular form of price determination in e-Commerce due to its simplicity and efficiency [2]. Auction is an attractive topic for the researchers to design a secure and practical protocol using cryptographic primitives.

The most common and basic types of auctions based on bidding mechanism are: the open English auction (ascending price), the open Dutch auction (descending price), the closed Sealed bid auctions (first or second price, the latter also known as Vickrey auction). By different classification criteria, the auction also can be divided into sequential and simultaneous auction, single-attribute and multi-attribute auction, single-object and multi-object auction, non-combinational and combinational auction, single auction and double auction and so on. According to the number of sellers and buyers, auctions can be divided into normal auction (one seller and many buyers) and reverse auction (many sellers and one buyer). Generally, B2C or C2C e-commerce auctions types often adopt normal auction, especially based on English auction like in eBay, while government procurement and companies purchasing through e-marketplace are often buyer-initiated reverse auctions.

Multi-attribute reverse auctions allow negotiation over non-price attributes such as quality, color, delivery time and so on [3].Multi-attribute auction is one of the rising and most beneficial procurement. Parkes et al. [5] scheme is extraordinary, which used advanced techniques and most secured protocol for price only and multi-item auctions. We use the same protocols of [5] and extended the work to multi-attribute auctions. We worked on designing a scoring function, which is useful in finding winner in multi-attribute auctions and helped us to extend the [5] scheme. In addition to that we added to use [14] work in generating pseudonyms which make bidders anonymous during the auction.

Rest of the paper is organized as follows. We explain related work in section 2 and describe the proposed protocol, phases involved in this scheme in section 3, security requirements and analysis in section 4, Implementation and results in section 5 and finally present our conclusions in section 6.

## 2. RELATED WORK

To the best of my knowledge there are only few papers dealing with security in multi-attribute auctions. Our target is to develop a multi-attribute reverse auction scheme with main focus on public verifiability and anonymity. Parkes et al. [5] scheme is efficient secure protocol for price only and multi-item auctions, which motivated us to extend this work to multi-attribute auctions. Yoones and Fathian [6] scheme is real time efficient work, in which many security features are achieved but they didn't deal with public verifiability. In this they clearly explained comparison of this protocol with their previous proposed protocols (EOBA (Electronic Online Bidding Auction) and SAP (Secure Auction Protocol) [7]). Rahman and Islam [8] scheme deals in constructing a trusted auction house. In our view protocol is not clear, no price flexibility and complex communications. Strecker and Seifert [9] experiment results illustrates that full revelation of the buyer's preferences significantly increases allocation efficiency. Hence in our method we reveal the preferences of auctioneer like scoring function, mechanism of selecting winner for increasing efficiency of scheme. Suzuki and Yokoo [10] scheme is the prime one in dealing with security in multi-attribute auction. This scheme applied to Vickrey auction and concentrated on bid privacy, public verifiability but there is no price flexibility.

## 3. PROPOSED PROTOCOL

The following protocol is proposed on basis of the scheme [5] with slight modifications for multi-attribute reverse auction. These modifications and protocol flow are explained clearly in this section. Overview in Figure 1 and its explanation are as follows:

1. Auctioneer publishes the terms of auction, the item, attributes and weights, timing constraints, auction id in bulletin board.
2. Bidders generate their unique pseudonym using generation algorithm in PDAs.
3. Bidders register for auction specified by auctioneer.
4. Auctioneer publishes Paillier encryption key and scoring function in bulletin board.
5. Bidders encrypt their attributes, generates score and its cipher. Ciphers of attributes are sent to auctioneer are accepted only within timing constraints specified.
6. Auctioneer publishes ciphers of all bidders. At the end of specified time he decrypts all ciphers to generate scores and its cipher.
7. Auctioneer stores all details in database server. Using Testsets and Range protocol he generates the hand over values, which are required for anyone to verify the auction outcome without revealing exact values.
8. Finally auctioneer publishes the hand over values in the bulletin board. Using hand over values in range protocol anyone can verify the auction outcome.

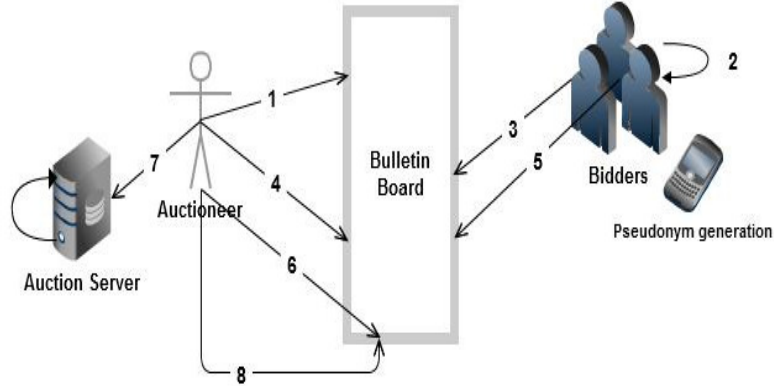

Figure 1. Protocol Overview

Protocol includes four phases: Initialization Phase, Bidding Phase, Opening and proof generation Phase, Winner announcement and Verification Phase. In our protocol, there are two participants, Buyer (auctioneer), and Bidder (sellers) and we use Paillier cryptosystem, Public Key Infrastructure (PKI), Certified Bulletin Board as explained in scheme [5].The following protocol is proposed on basis of the scheme [5] with modifications for multi-attribute reverse auction. These modifications and protocol flow are explained clearly in this section. Protocol includes four phases: Initialization Phase, Bidding Phase, Opening and proof generation Phase, Winner announcement and Verification Phase. In our protocol, there are two participants, Buyer (auctioneer), and Bidder (sellers) and we use Paillier cryptosystem, Public Key Infrastructure (PKI), Certified Bulletin Board as explained in scheme [5].

### 3.1. Initialization Phase

**Step 1:** Auctioneer posts the following information on the bulletin board: the terms of the auction specifying the item, Scoring function, the mechanism for selection of the winner in multi-attribute auction, the deadline T, an identifier ID of the auction, and a Paillier encryption key n. Auctioneer knows the corresponding decryption key. Paillier encryption and decryption and scoring function are explained as follows.

### 3.1.1. Paillier Cryptosystem

The Paillier cryptosystem [13], named after and invented by Pascal Paillier in 1999, is a probabilistic asymmetric algorithm for public key cryptography. It is a Trapdoor Discrete Logarithm Scheme. Paillier's trapdoor function is an isomorphism $f: Z_N \times Z^*_N \to Z^*_{N^2}$.

Where $Z_N$ is integers modulo N and $Z^*_N$ is the multiplicative group of integers modulo n, i.e. the set of integers relatively prime to n.

 Encryption

- $c = g^M r^n \mod n^2$ where M is plaintext (attribute values), c is ciphertext, $n = p \cdot q$ where p, q are odd primes, g is an integer belongs to $Z_N$, r is a random number in $Z_n^*$

Decryption

- $M = L(c^{\emptyset(n)} \mod n^2) \cdot \mu \mod n$, where $\mu = \varphi(n)^{-1} \mod n$

$L(u) = (u-1)/n$, $\emptyset(n) = \varphi(n) = (p-1)(q-1)$, size of $Z^*_n$, $\varphi$ is Euler's totient function

### 3.1.2. Scoring Function: Winner determination in Multi-attribute auctions

Winner determination is the process of selecting the right bidder to award the business. A buyer in multi-attribute reverse auction tries to minimize the price while maximizing the other quality related attributes during the winner determination. In other words she tries to maximize her utility defined through a scoring function [11] [12]. A scoring function [15] combines the attributes to determine a score for each multi-attribute bid. A typical scoring function takes the following form:

$$S = \sum_{r=1}^{K} w_r \cdot f_r(x_r) - P$$

Where, K is the number of non-price attributes considered by the buyer, $x_r$ is the value of the $r^{th}$ attribute submitted by the supplier, $w_r$ is the weight indicating the importance given to the attribute, $f_r(.)$ is the valuation function associated with the attribute, and P is the bid price. The summation term inside S is also called the aggregate value function. The buyer maximizes the scoring function to selecting the supplier.

**Step 2:** Auctioneer will sign the entire data posted on bulletin board.

**Step 3:** Bidders generates pseudonym using algorithm [14] and he registers in the auctioneer's site using this pseudonym.

### 3.2. Bidding Phase

**Step 1:** Every Bidder chooses attribute values mentioned by the auctioneer like price, quality, delivery time, quantity etc. Bidder will find their respective scores using the scoring function given by auctioneer in Initialization phase. In order to create efficient test sets to prove bid sizes, we restrict the size of the bid so that $bid_i < 2^t < n/2$ for small t.

**Step 2:** Bidders encrypts all the attribute values using the public key n (using Paillier algorithm) and a randomly chosen help value r $[C_i = E(bid_i, r_i)]$. They also encrypt score, which is useful during verification phase.

**Step 3:** Bidders then sends all ciphers to auctioneer. Finally, the bidder sign all the information sends to auctioneer before time T. Auctioneer returns a signed receipt to Bidder.

**Step 4:** After time T Auctioneer posts the encrypted bids in bulletin board. Also signs and posts the Testsets for verification of limit of the bid value using range protocol, which will be explained in next phase.

**Step 5:** Bidder who has a receipt for a bid which is not posted, can appeal her non-inclusion.

### 3.3. Opening and proof generation Phase

**Step 1:** Auctioneer decrypts all the bids using Paillier decryption key and computes the auction result that is score using scoring function. Also auctioneer calculates associated random 'r' values used for encrypting the bid for constructing the Proof of correctness.

**Step 2:** Auctioneer computes the score values of all bidders and verify the score sent by them. He then finds the winner with maximum score. Auctioneer (uses ciphers of score instead of price in [5] generates hand over values using inequality proof and Testsets for proving the determined winner. Handover values are values which are used by anyone to verify the winner published. From scheme [5] Range protocol, test sets and hand over values are explained in properties of Paillier cryptosystem as follows.

### 3.3.1. Properties

1. Homomorphic: $E(M_1 + M_2) = E(M_1) \times E(M_2)$, $E(k \times M) = E(M)^k$

2. Self-blinding: $D(E(M) r^n \mod n^2) = M$

3. Inequality comparison:

    a. Given two ciphertexts $C_x = E(x)$ and $C_y = E(y)$; Prover can show $x > y$ and $x \geq y$. Because our values x and y are integers mod $n^2$, we can prove $x > y$ by showing $x \geq y + 1$, provided $y \neq n - 1$.

    b. Due to the homomorphic properties of Paillier encryption $E(x+1) = E(x) \times (n+1)$ (mod $n^2$), and so adding 1 to a value in its encrypted form is trivial. Thus, all ordering comparisons can be reduced to the ability to prove $x \geq y$.

    c. x and y must be in the range $[0, 2^t)$ for $2^t < n/2$.

    d. To prove $x \geq y$ both prover and verifier calculate $E(x-y) = E(x) \times E(y)^{-1} \mod n^2$ and P proves $0 \leq x-y < 2^t < n/2$ from $E(x-y)$ using Test set and Range protocol.

    **Test Set definition:** A valid test set TS for the assertion ''$C = E(x,r)$ is an encryption of a number $x < 2^t < n/2$'' is a set of 2t encryptions $TS = \{G_1 = E(u_1, s_1), \ldots, G_{2t} = E(u_{2t}, s_{2t})\}$, where each of the powers of $2:1, 2, \ldots, 2^{t-1}$ appears among the $u_i$ exactly once and the remaining t values $u_j$ are all 0. Each test set's elements are randomly ordered. By use of test set TS, the prover can prove that $x < 2^t < n/2$ using range protocol.

    **Range Protocol:** Let $x = 2^{t1} + \ldots + 2^{tl}$ be the representation of x, a sum of distinct powers of 2. Auctioneer selects TS from encryptions $G_{j1}, \ldots, G_{jl}$ of $2^{t1} + \ldots + 2^{tl}$ and further t-l $G_{jt+1}, \ldots, G_t$ encryptions of 0.

    $(E(x, r)^{-1} \cdot G_{j1} \ldots G_{jt}) \pmod{n^2} = E(0, s)$ …………………………I

    R.H.S is an encryption of 0 with help values $s = (r^{-1} \cdot s_{j1} \ldots s_{jt}) \pmod n$ if and only if indeed $x = 2^{t1} + \ldots + 2^{tl}$. Hence auctioneer has decryption key can know the value r, then he can hand over the set $\{G_{j1} \ldots G_{jt}\}$ and help value 's'. Anyone can verify on their own that above equation holds and deduce that $x < 2^t < n/2$.

    This protocol reveals nothing to verifier beyond $x < 2^t < n/2$ because TS is a set of elements and randomly permutated and no information about the values. Consequently has no information about which encryptions of powers of 2 are included in $\{G_{j1} \ldots G_{jt}\}$. Furthermore, the inclusions of t –l encryptions of 0 hides even the number of non-zero bits in the binary representation of x. Finally, the random factors $s_{j1}, \ldots, s_{jt}$ present in the test set's encryptions combine to a uniformly random s, which completely masks any information about the help value r in the encryption $E(x, r)$. Consequently no information about x is revealed.

### 3.4. Winner announcement and Verification Phase

**Step 1:** Auctioneer publishes Winner score, his pseudonym and hand over values. Auctioneer proves the winner computed without revealing the exact bided values.

**Step 2:** Anyone can verify the result the using the test sets and hand over values in range protocol (as explained in Opening and proof generation Phase) posted by auctioneer on bulletin board.

## 4. SECURITY ANALYSIS

The security requirements [4] for sealed-bid auction protocol are examined as follows.

1. **Correctness:** All party acts honestly the correct wining price and winner is determined according to auction rules.
2. **Confidentiality:** No bidder knows anything other bidders bid before submits his own bid. This can be achieved using Paillier's probabilistic encryption. For one message we will get different ciphertexts none can guess messgage using ciphertext.
3. **Bidder-Anonymity:** The bidder's identity cannot be disclosed to any body throughout the auction. This is achieved using unique user generated pseudonym algorithm. Unless and until bidder discloses his identity, auctioneer is not able to know the bidder.
4. **Non repudiation:** After a bidder submits his bid, the bid cannot be modified. No bidder can deny his bid after he submits it. This is achieved by sign on all the information posted in bulletin board.
5. **Bid-Privacy:** The losing bids are not revealed until end of the auction even to the auctioneer. This scheme supports weak privacy. Bids are revealed to auctioneer but cannot collude with anyone because he doesn't have any information except the pseudonym of bidder.
6. **Public verifiability:** Anybody can verify the auction outcome. Using Paillier homomorphic properties and zero knowledge proof as explained in protocol phases this can be achieved.
7. **Price flexibility:** Bidder able to bid any value within a range. Bidder can bid any value within a range, less than $2^t$).

## 5. RESULTS AND ANALYSIS

We implemented the public verifiability in multi-attribute auctions using scoring function and for anonymity we simulated the pseudonym generation using algorithm in [14]. The score generated using attributes values bided by bidders is used place of price in [5]. We implemented the protocols in java using BigInteger in java.Math and Server Configuration: Microsoft Windows Server 2003 R2, Standard edition Service pack 2, Intel® Xeon™,CPU 3.60 GHz, 3.12 GB of RAM. Our results are quite good and took less time. By analyzing results we can say that it can be easily deployed in real time for multi-attribute auctions public verifiability. Summary of our results are in table 1 and 2 respectively.

Table 1. Proof Preparation.

| Key size (in bits) | Number of bids | Time(in milliseconds) |
|---|---|---|
| 512 | 100 | 954266 |
| 512 | 200 | 7115140 |
| 512 | 300 | 8610937 |
| 1024 | 100 | 7115140 |
| 1024 | 200 | 27897281 |
| | | |

Table 2. Verification.

| Key size (in bits) | Number of bids | Time(in milliseconds) |
|---|---|---|
| 512 | 100 | 56010 |
| 512 | 200 | 232672 |
| 512 | 300 | 541407 |
| 1024 | 100 | 1289906 |
| 1024 | 200 | 1475297 |

## 6. CONCLUSION

The proposed scheme aims in secure multi-attribute reverse auction which focuses mainly on public verifiability and anonymity in multi-attribute auctions. Our work includes public verifiability and anonymity in multi-attribute auctions are implemented using Parkes et al [5] scheme and Peter's [14] unique user generated digital pseudonym algorithm. Highlight of the scheme are using scoring function and without revealing the exact values public verifiability is achieved. This scheme is not complex or costly because no trusted parties and no distributed computation involved. This scheme is simple, secure enough, easy to implement and efficient as per our results and analysis.

## ACKNOWLEDGEMENTS

We would like to thank Information security Labs, Computer Science and Engineering Department, NITK, Surathkal.

**Authors**

**T. R. Srinath** received his Bachelor of Engineering degree in Computer Science and Engineering from Maturi Venkata Subba Rao Engineering College, Osmania University Hyderabad, Andhra Pradesh. He was born in Andhra Pradesh, INDIA. He is currently pursuing his Master of Technology Degree in Computer Science & Engineering-Information Security from NITK, Surathkal. His areas of interest are Information Security, Cryptography and Cloud computing. His email is talari.srinath@gmail.com.

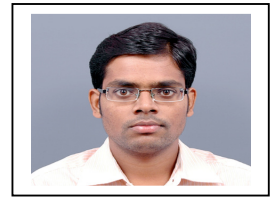

**Mahendra Pratap Singh** is currently the Assistant Professor, Department of Computer Science and Engineering, National Institute of Technology Karnataka, Surathkal, INDIA. He was born in Uttar Pradesh, India. He completed his Bachelor of Engineering from the Uttar Pradesh Technical University, Uttar Pradesh and Master of Engineering from Karunya University, Coimbatore (TN). His areas of interest are Information Security, Cryptography, Image Processing, Parallel and Distributed Computing and Wireless Network.

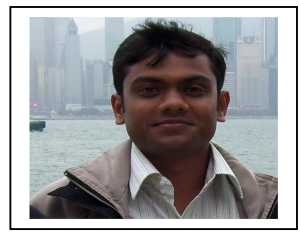

**Alwyn R. Pais** is currently the Assistant Professor, Department of Computer Science and Engineering, National Institute of Technology Karnataka, Surathkal, INDIA. He was born in Karnataka, India. He completed his Bachelor of Engineering from the Mangalore University, Karnataka and Master of Technology from IIT Bombay. He is currently pursuing his PhD from NITK, Surathkal. He is also the coordinator of the ISEA project at NITK. His areas of interest are Algorithms, Cryptography and Computer Vision. His email is alwyn.pais@gmail.com and has personal webpage at http://isea.nitk.ac.in/faculty/alwyn/

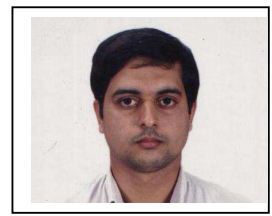